\documentclass{naturefig}
\usepackage{graphicx}

\newcommand{\gtappeq}{\raisebox{-0.6ex}{$\,\stackrel{\raisebox{-.2ex}{$\textstyle >$}}{\sim}\,$}}

\newcommand{\msun}{M_\odot}
\newcommand{\rsun}{R_\odot}

\def\spose#1{\hbox to 0pt{#1\hss}}
\def\lta{\mathrel{\spose{\lower 3pt\hbox{$\mathchar"218$}} \raise 2.0pt\hbox{$\mathchar"13C$}}}
\def\gta{\mathrel{\spose{\lower 3pt\hbox{$\mathchar"218$}} \raise 2.0pt\hbox{$\mathchar"13E$}}}
\newcommand{\aplett}{\textit{Astrophys. Lett.}}
\newcommand{\apj}{\textit{Astrophys. J.}}
\newcommand{\aj}{\textit{Astronom. J.}}
\newcommand{\apjl}{\textit{Astrophys. J. Lett.}}
\newcommand{\apjs}{\textit{Astrophys. J. Supp.}}
\newcommand{\mnras}{\textit{Mon. Not. R. Astron. Soc.}}

\newcommand{\aap}{\textit{Astron. \& Astronphys.}}

\title{\Large \bf A Binary Origin for Blue Stragglers in Globular Clusters}

\author{Christian Knigge$^{1}$, Nathan Leigh$^2$, \& Alison Sills$^2$}

\begin{document}

\topmargin=0.7in

\spacing{1.0}
\maketitle

\begin{affiliations}
 \item University of Southampton, School of Physics and Astronomy,
   Southampton SO17 1BJ, UK 
 \item McMaster University, Department of Physics and Astronomy, 1280
   Main Street West, Hamilton, ON, L8S 4M1, Canada 
 \end{affiliations}

\begin{abstract}

Blue stragglers in globular clusters are abnormally massive stars that
should have evolved off the stellar main sequence long ago. There are
two known processes that can create these objects: direct stellar
collisions\cite{1976ApL....17...87H} or binary
evolution\cite{1964MNRAS.128..147M}. However, the relative importances
of these processes have remained unclear.
 In particular, the total
number of blue stragglers found in a given cluster does not seem to
correlate with the predicted collision
rate\cite{2004ApJ...604L.109P,2007ApJ...661..210L},
providing indirect support for the binary scenario. Yet the radial
distributions of blue stragglers in many clusters are bimodal, with a
dominant central
peak\cite{1993AJ....106.2324F,2004ApJ...603..127F,2006MNRAS.373..361M}. 
This has been interpreted as an
indication that collisions do dominate blue straggler production, at
least in the high density cluster cores\cite{2006MNRAS.373..361M, 2004ApJ...605L..29M}.  
Here, we show that there is a clear, but sub-linear, correlation between the number 
of blue 
stragglers found in a cluster core and the total stellar mass
contained within it. This is the strongest and most direct evidence to
date that most blue stragglers, even those found in cluster cores, are
the progeny of binary systems. However, we also point out that the
parent binaries may themselves have been affected by dynamical
encounters. This may be the key to reconciling all of the seemingly
conflicting results found to date.  

\end{abstract}

In both blue straggler (BS) formation scenarios, the numbers and
properties of the resultant BSs can shed light on the dynamical
history of their host clusters. The conclusions drawn from the
observed BS populations depend on which formation mechanism is
dominant. Unfortunately, even though a number of potential formation
tracers have been proposed, a simple observational distinction between
collisional and binary BSs has yet to be found
\cite{1987ApJ...323..614B,1995ApJ...447L.121L,1995ApJ...455L.163P,1997ApJ...477.
  .335S,1999ApJ...513..428S,2006ApJ...647L..53F}. In general, it is
impossible to ascertain whether an individual BS was formed via a
stellar collision or a binary merger.

However, it should be possible to determine the dominant formation
channel statistically, by studying the relationship between the size
of a BS population and the properties of its host cluster. Here, we
focus on BSs found in the cores of clusters, where
collisions should be most frequent. If most BSs are formed as a result
of single-single collisions, their number in a particular cluster core
should be approximately $N_{BSS,coll} \simeq \tau_{BSS} /
\tau_{coll}$, where $\tau_{BSS}$\ is the typical BS lifetime, and
$\tau_{coll}$ is the time-scale between single-single stellar
collisions in the core. Conversely, if most core BSs are the progeny
of binary stars, their number may be expected to scale simply as
$N_{BSS,bin} \propto f_{bin} M_{core}$, where $f_{bin}$\ is the binary
fraction in the core and $M_{core}$\ is the total stellar mass
contained in the core. These predictions are for the simplest versions
of the competing scenarios, but note that more complex, intermediate channels
could also contribute (e.g. BSs produced as a result of dynamical
encounters involving binary stars). We will return to this subtlety 
below.

In order to test these simple predictions, we used our existing data
base of core BS counts
\cite{2007ApJ...661..210L},
which are derived from a large set of {\em HST}-based colour-magnitude
diagrams\cite{2002A&A...391..945P}. Figure~1 confirms that there is no
global correlation between the observed core BS numbers and those
predicted by the collision 
scenario\cite{2004ApJ...604L.109P,2007ApJ...661..210L}.
\nocite{1986ApJ...303..336G}
\nocite{1989AJ.....98..217L} 
\nocite{1996AJ....112.1487H}
\nocite{2005ApJS..161..304M}
\nocite{1993ASPC...50..357P}
\nocite{1985IAUS..113..541W}
\nocite{1997ApJ...487..290S,2001ApJ...548..323S}
In particular, the predicted numbers are very different for high- and
low-density clusters, which is in conflict with the observations. By
contrast, Figure~2 shows that the observed numbers do correlate
strongly with the cluster core masses, as expected if most BSs are the
progeny of binary stars. A power law fit suggests $N_{BSS} \propto
M^{0.38 \pm 0.04}$.
\nocite{1986ApJ...303..336G}
\nocite{2007HiA....14..440V}

When consideration is restricted to dense cores
($\rho_0 \gtappeq 10^4 ~ {\rm M_{\odot} ~ pc}^{-3}$), the collisional
predictions are, in fact, significantly correlated with  
observed BS numbers (see the black points in Figure~1). However, even
for these dense cores, the correlation with core mass remains stronger
(Figure~2). Thus not only is core mass the only successful global 
predictor for core BS numbers, but it remains a better predictor than 
collision rate even for dense clusters. 


One way to distinguish even more clearly between collisional and
binary scenarios is to isolate the key cluster parameters that drive
the observed correlations and quantify their impact on the observed
sizes of the core BS populations. For example, even though collision
rate and core mass are correlated parameters, only collision rate
depends fundamentally on a cluster's central velocity dispersion. We therefore
carried out a direct fit to the observed core BS numbers, using the
generalized model $N_{BSS} = K \; r_c^{\alpha} \; \rho_0^{\beta} \;
\sigma_0^{\gamma}$, where $K$, $\alpha$, $\beta$\ and $\gamma$ are all
treated as free parameters.  For purely collisional BS formation, we
expect $\alpha \simeq 3$, $\beta \simeq 2$ and $\gamma \simeq -1$,
whereas pure binary formation implies $\alpha \simeq 3$, $\beta \simeq
1$ and $\gamma \simeq 0$. These relationships assume constant 
binary fraction across clusters; we will return to this assumption below. 
Allowing for a more general power-law dependence on collision rate or 
core mass, we 
predict $\alpha \simeq 1.5 \beta \simeq -3 \gamma$\ for collisional BS
formation, and $\alpha \simeq 3 \beta$\ with $\gamma \simeq 0$\ for
the binary scenario.

Figure~3 shows the results of this fit. 
\nocite{1986ApJ...303..336G}
The best-fit power law indices
are $\alpha = 1.47 \pm 0.29$, $\beta = 0.55 \pm 0.15$ and $\gamma =
-0.40 \pm 0.27$. These coefficients are consistent with a
power-law dependence on $M_{core}$, but not with one on collision
rate. (Note that since $r_c$, $\rho_0$\ and $\sigma$\ are correlated 
parameters, $\alpha$, $\beta$\ and $\gamma$\ are also correlated.)
There is no evidence for a significant dependence of
BS numbers on velocity dispersion. A fit to the dense clusters only
yielded results which were consistent with the values obtained for the
complete cluster sample.  

We find that BS numbers depend strongly on core mass, but not on
collision rate. We argue that this is strong evidence for a binary, as
opposed to a collision, origin for blue stragglers in globular
clusters.  We note that the core mass dependence can be approximated
as $N_{BSS} \propto M_{core}^{\delta}$\ with $\delta \simeq 0.4 -
0.5$.  The sub-linear dependence of BS numbers on core mass is
interesting. For ordinary populations of single stars, one would 
expect a roughly linear dependence, and we have verified that this is
the case for the red giant and horizontal branch populations in our
cluster sample. In the context of the binary scenario for BSs, the
simplest way to account for this sub-linearity is to posit that core
binary fractions decrease with increasing core mass, i.e. $f_{bin}
\propto M_{core}^{-\epsilon}$ with $\epsilon \simeq 0.5 - 0.6$. This
is a testable prediction. There have been two recent efforts to
estimate core binary fractions in a consistent way from homogeneous,
deep HST/ACS observations of globular
clusters\cite{2007MNRAS.380..781S,2008arXiv0801.3177M}. One of
these\cite{2007MNRAS.380..781S} is restricted to a set of 13
low-density clusters (Sample 1). The other\cite{2008arXiv0801.3177M}
is still preliminary, but based on a larger set of 35 clusters that
span a wider range in density and other dynamical parameters (Sample
2). Promisingly, from our perspective, a correlation between BS
frequency and core binary fraction has already been reported for
Sample 1\cite{2008A&A...481..701S}, and an anti-correlation between
core binary fraction and total cluster mass has been reported for
Sample 2\cite{2008arXiv0801.3177M}.  Since Sample 1 spans only a
narrow dynamic range, and the binary fractions reported for Sample 2
are only available in graphical form, we could only make a preliminary
attempt to infer the dependence of $f_{bin}$\ on $M_{core}$\ from
these samples. Nevertheless, our power law estimate is
$f_{bin} \propto M_{core}^{-0.35}$\ for both samples, close to the
predicted relation, albeit with considerable scatter. Thus the
number of BSs found in the cores of globular clusters scales roughly 
as $N_{BSS} \propto f_{bin} M_{core}$, just as expected if most core
BSs are descended from binary systems. 

There are several ways to build on the analysis presented above.
First, binary-single and binary-binary interactions may actually
produce more 
blue stragglers in clusters than single-single collisions because the
interaction cross sections are larger. The expected number of
binary-single [binary-binary] collisions scales like the number of
single-single 
collisions times $f_b$ [$f_b^2$]. A better determination of the
dependence of binary fractions on cluster parameters is clearly
warranted to test these predictions fully. However, based on the 
preliminary analysis of the observed binary fractions described above, 
it currrently seems unlikely that binary-single and binary-binary
interaction numbers can produce the observed scaling of blue stragger
numbers with cluster parameters. 

Second, the effects of mass segregation have been ignored in our
analytical predictions. Massive stars at any particular location in a
cluster will sink to the core on roughly the local relaxation time.
Thus some of the BSs found today in cluster cores may actually have
been formed outside the
core\cite{2004ApJ...605L..29M,2006MNRAS.373..361M,2004MNRAS.349..129D}.
Moreover, clusters evolve, so the present-day cluster parameters may
not always be close approximations to what should ideally be
suitably-weighted averages over the BS lifetime. As a first test, we
have split our cluster sample in objects with half-mass relaxation times
smaller and greater than the blue straggler lifetime. If mass
segregation were responsible for masking an underlying trend with
collision rate, we may expect the long relaxation timescale clusters 
to reveal this trend, while the short relaxation time clusters should
exhibit enhanced BS numbers (at a given collision rate). We find no
evidence for either of these effects. It is also worth noting that our result
for core BSs is actually consistent with the behaviour of the total
(core+peripheral) BS populations, which also seem to show a
sub-linear correlation with total cluster mass \cite{2004MNRAS.349..129D}.

Our results do not imply that dynamical interactions are
unimportant to BS formation. The properties of the core binary
populations from which BSs are probably descended may themselves be
strongly affected by dynamical encounters. There is a wide spectrum
of possible paths within the family of binary channels. While some
'binary BSs' may be the direct descendants of primordial binaries that
evolved essentially in isolation, others may exist only as a result of
multiple strong interactions and exchanges
\cite{2006ApJ...641..281K,2008arXiv0805.0140K}. We are not the first
to point out this subtlety (though it has sometimes been obscured by
the terminology used). For example, Mapelli et al. note that the
collisional BS population invoked in their model could be produced via
encounters involving
binaries\cite{2004ApJ...605L..29M,2006MNRAS.373..361M}, and the
primordial binary channel described by Davies et al. explicitly allows
for dynamical interactions\cite{2004MNRAS.349..129D}. Our results do, 
however, argue against the simple scaling of blue straggler
numbers with any collision rate. Clearly, a key
challenge for future work will be to determine if dynamically active
binaries can produce BS numbers consistent with the trends found here.

\bibliographystyle{naturemag.bst}

\begin{addendum}
 \item[Acknowledgements] Research support for C.K. was provided by the UK's Science and
   Technology Facilities Council. A.S. and N.L. are supported by the
   Natural Sciences and Engineering Research Council of
   Canada. C.K. would like to thank Tom Maccarone for useful
   discussions.

\item[Author Information]Reprints and permissions information is
  available at \\ npg.nature.com/reprintsandpermissions. The authors
  declare that they have no competing financial interests. Correspondence
  and requests for materials should be addressed to C. K. ~(email:
  christian@astro.soton.ac.uk).
\end{addendum}

\clearpage

\begin{figure}
\begin{center}
\includegraphics[scale=0.6, angle=270]{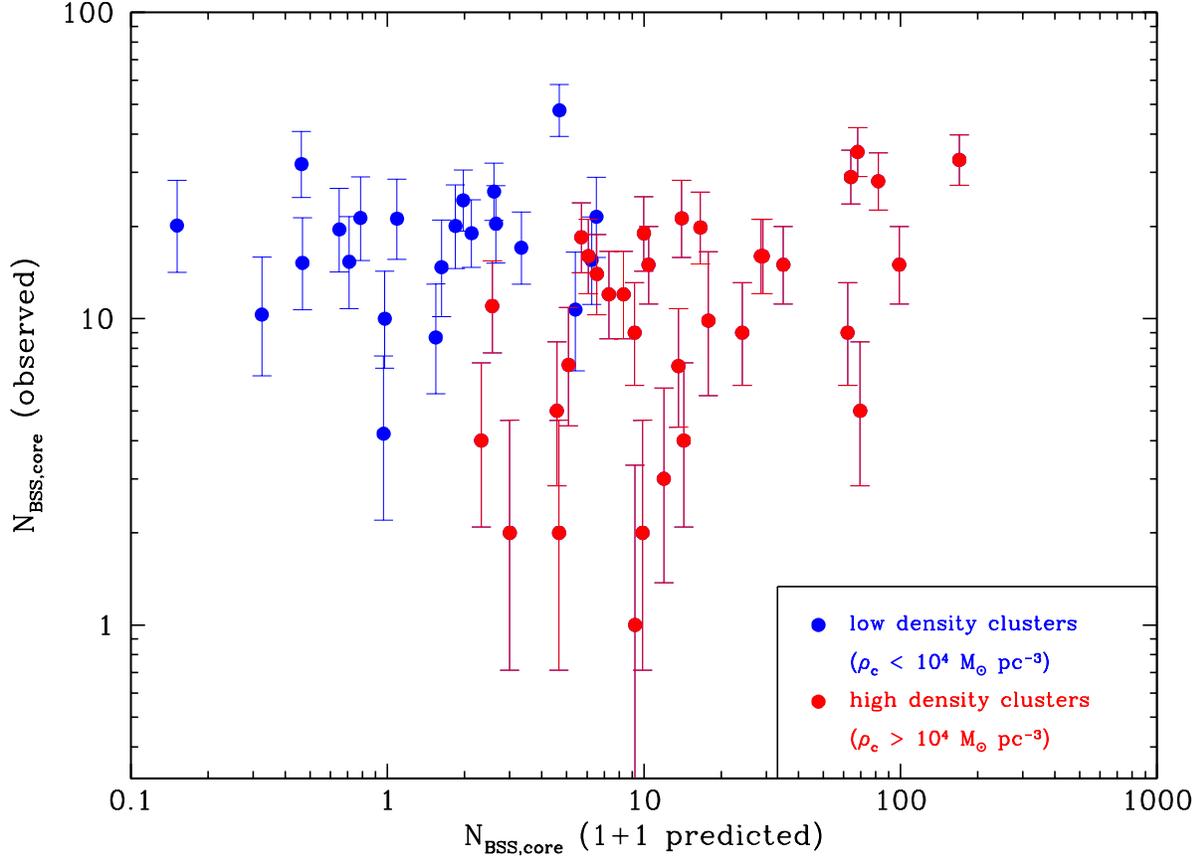}
\end{center}
\caption[]{The observed numbers of core blue stragglers versus the numbers
expected from single-single collisions. Black points correspond to
high-density clusters ($\rho_0 > 10^4 ~ {\rm M_{\odot} ~ pc}^{-3}$), 
grey points to
low-density ones ($\rho_0 < 10^4 ~ {\rm M_{\odot} ~ pc}^{-3}$). 
The error bars shown
are approximate 68\%  Poisson confidence
intervals\protect\cite{1986ApJ...303..336G}. There is no
correlation betwen observed and predicted numbers across the full
sample (Spearman's rank test yields $\rho_s = -0.03$, $p = 0.6$, $N =
56$). For the subset of dense clusters, a correlation does
exist ($\rho_s = 0.52$, $p = 8 \times 10^{-4}$, $N = 34$). 
In calculating the predicted numbers, we took the 
time-scale between collisions to be\protect\cite{1989AJ.....98..217L} 
\protect\begin{math}
\tau_{coll} = 1.1 \times 10^{10}
\left(\frac{r_c}{1~{\rm pc}}\right)^{-3}
\left(\frac{n_0}{10^3~{\rm pc^{-3}}}\right)^{-2}
\left(\frac{\sigma_0}{5~{\rm km~s^{-1}}}\right)
\left(\frac{M_*}{\msun}\right)^{-1}
\left(\frac{R_*}{\rsun}\right)^{-1}
{\rm years.}
\protect\end{math}
Here $r_c$\ is the core radius, $n_0$\ is the central number density
in stars per cubic parsec, $\sigma_0$\ is the central velocity
dispersion, $M_*$\ is the average stellar mass (taken to be 0.5
$\msun$), and $R_*$\ is the average stellar radius (0.5 $\rsun$). Core
radii and central luminosity densities were taken from the McMaster
Globular Cluster Catalogue\cite{1996AJ....112.1487H}. Core number
densities were estimated by adopting a fixed mass-to-light ratio of
2\cite{2005ApJS..161..304M}. Central velocity dispersions were taken
primarily from the compilation of Pryor \&
Meylan\cite{1993ASPC...50..357P} , but supplemented with data from
Webbink\cite{1985IAUS..113..541W}. The typical BS lifetime in a
cluster was taken to be 1.5 Gyr
\cite{1997ApJ...487..290S,2001ApJ...548..323S}. Note that for
high-concentration clusters, the observational definition of the core
radius can be somewhat arbitrary. However, this should not have a
significant impact on our analysis, as long as the associated central
densities and velocity dispersions remain reasonable averages over the
relevant regions.}
\label{fig1}
\end{figure}

\thispagestyle{empty}
\begin{figure}
\begin{center}
\includegraphics[scale=0.6, angle=270]{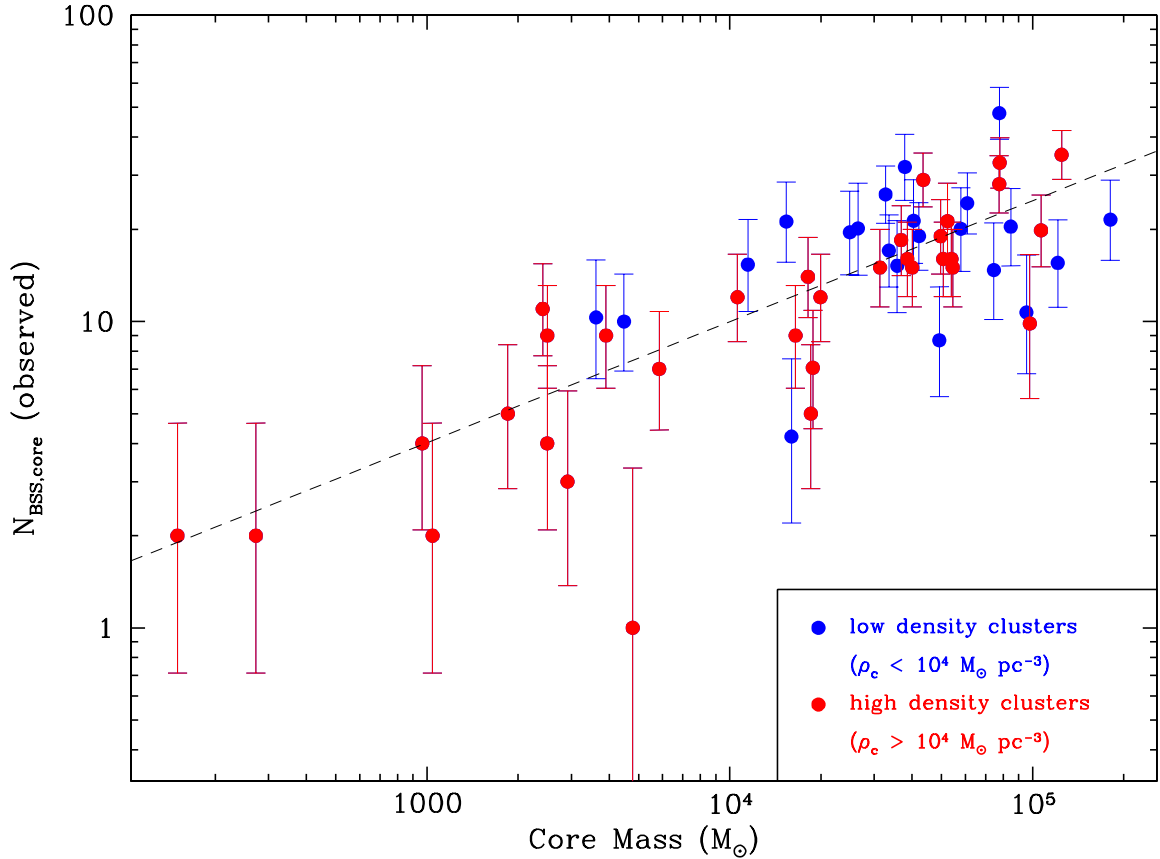}
\end{center}
\caption[]{The observed numbers of core blue stragglers versus the
estimated core masses. Black points correspond to
high density clusters ($\rho_0 > 10^4 ~ {\rm M_{\odot} ~ pc}^{-3}$), 
grey points to
low-density ones ($\rho_0 < 10^4 ~ {\rm M_{\odot} ~ pc}^{-3}$). 
The error bars shown
are approximate 68\%  Poisson confidence
intervals\protect\cite{1986ApJ...303..336G}. BS numbers and core
masses are strongly correlated ($\rho_s = 0.71$, $p = 4 \times
10^{-10}$, $N = 56$), as expected if BSs are descended from binary
systems. The correlation with core mass also holds for the subset of
dense clusters ($\rho_s = 0.84$, $p = 3 \times 10^{-10}$, $N = 34$) and
in both cases is stronger than that with collision rate. The dashed
line is a power law fit to the full set of clusters, 
obtained with a method appropriate for this type of data
set\protect\cite{2007HiA....14..440V}. The best-fitting power law is $N_{BSS}
\propto M^{0.38 \pm 0.04}$.}
\label{fig2}
\end{figure}

\thispagestyle{empty}
\begin{figure}
\begin{center}
\includegraphics[scale=0.6,angle=270]{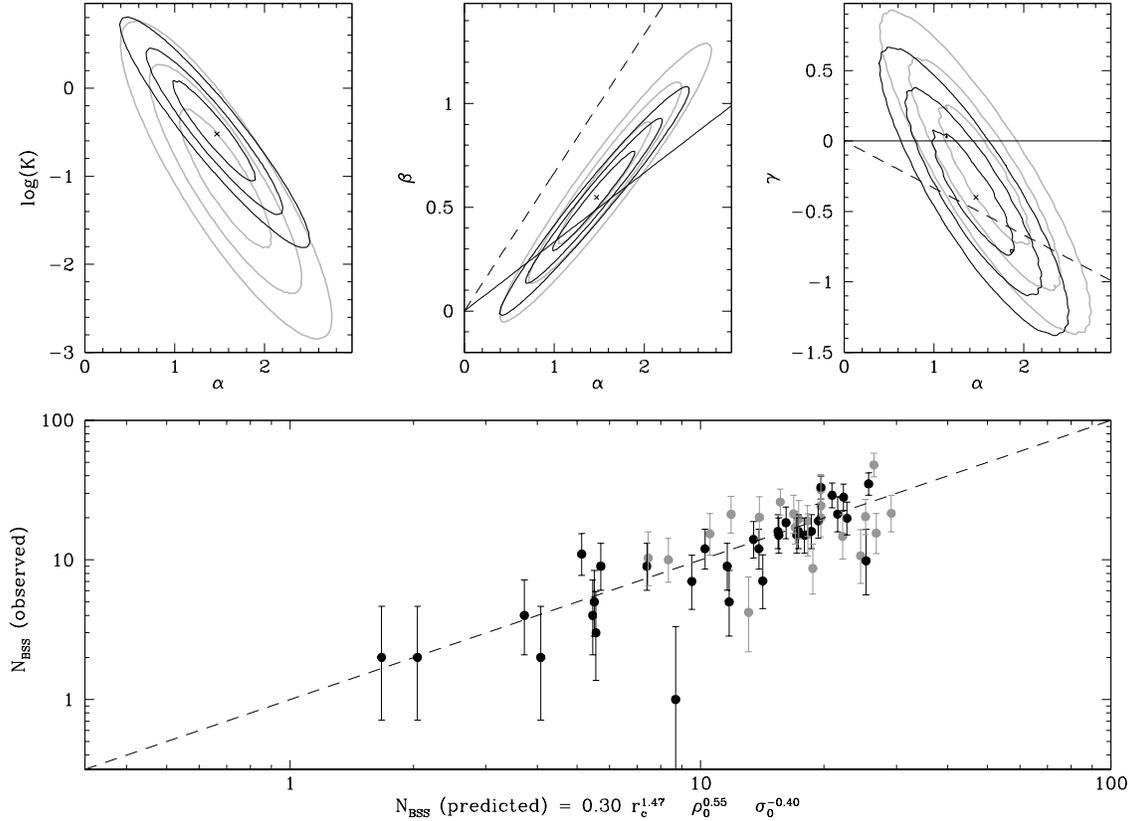}
\end{center}
\caption[]{Results of fitting the generalized model $N_{BSS} \propto K
  \; r_c^{\alpha} \; \rho_0^{\beta} \; \sigma_0^{\gamma}$\ to the
  observed BS numbers. The top panels show three projections of the
  4-dimensional parameter space. The contours in each panel represent
  1$\sigma$, 2$\sigma$ and 3$\sigma$ joint confidence intervals on the
  two parameters being considered. Black contours correspond to the
  full cluster sample, grey contours to the dense cluster sample ($\rho_0
  > 10^4 ~ {\rm M_{\odot} ~ pc}^{-3}$). The straight lines in the 
  middle and right
  panels show the parameter relationships expected for a power law
  dependence on collision rate (dashed lines) and core mass (solid
  line). The best fit power law indices are $\alpha = 1.47 \pm 0.29$,
  $\beta = 0.55 \pm 0.15$ and $\gamma = -0.40 \pm 0.27$.  The bottom
  panel shows observed BS numbers vs the best-fit model prediction for
  the full sample. Black points correspond to high density clusters
  ($\rho_0 > 10^4 ~ {\rm M_{\odot} ~ pc}^{-3}$), 
  grey points to low-density ones
  ($\rho_0 < 10^4 ~ {\rm M_{\odot} ~ pc}^{-3}$). 
  The error bars shown are approximate
  68\% Poisson confidence intervals\protect\cite{1986ApJ...303..336G}.
  The dashed line marks the identity relation, i.e. $N_{BSS}(observed)
  = N_{BSS} (predicted)$.} We also carried out a separate fit to the
subset of dense clusters ($\rho_0 > 10^4~{\rm M_{\odot} ~ pc}^{-3}$). 
This yielded
$\alpha = 1.59 \pm 0.33$, $\beta = 0.63 \pm 0.19$ and $\gamma = -0.28
\pm 0.31$, consistent with the values obtained for the complete
cluster sample. Note that, even within this subset, there is no
evidence for a dependence of BS numbers on $\sigma_0$.
\label{fig3}
\end{figure}

\end{document}